# Biometric Blockchain: A Secure Solution for Intelligent Vehicle Data Sharing

Bing Xu[1], Tobechukwu Agbele [2], Qiang Ni[3] and Richard Jiang[4]

**Abstract –** The intelligent vehicle (IV) has become a promising technology that could revolutionize our life in smart cities sooner or later. However, it yet suffers from many security vulnerabilities. Traditional security methods are incapable to secure the IV data sharing against malicious attacks. Blockchain, as expected by both research and industry communities, has emerged as a good solution to address these issues. The major issues in IV data sharing are trust, data accuracy and reliability of data sharing in the communication channel. Blockchain technology, previously working for the cryptocurrency, has recently applied to build trust and reliability in peer-to-peer networks with similar topologies of IV data sharing. In this chapter, we present a new framework, namely biometric blockchain (BBC), for secure IV data sharing. In our new scheme, biometric information is exploited as a cue to record who is responsible in the data sharing activities, while the proposed BBC technology serves as the backbone of the IV data-sharing architecture. Hence, the proposed BBC technology provides a more reliable trust environment between the vehicles while personal identities are traceable in the proposed new scheme.

## 1. Introduction

Intelligent vehicles (IVs) [1-5] are empowered by its capability of internet access with real-time data processing and sharing over *Vehicular Ad-hoc Networks* (VANET), where moving IVs are dynamically integrated into pervasive networks in smart cities. Within this context, IV data sharing faces the security challenges in two typical scenarios: Vehicle-to-Vehicle (V2V) data sharing and Vehicle-to-Infrastructure (V2I) data sharing [1]. Consequently, a secure protocol is needed to provide the regulation of any safety-critical incidents and hazards.

[1] Computer & Information Sciences, Northumbria University, Newcastle upon Tyne, NE1 8ST, UK, e-mail: bing.xu@northumbria.ac.uk.

[2] Computing & Communication, Lancaster University, Lancaster, LA1 4WY, UK, e-mail: t.agbele@lancaster.ac.uk.

[3] Computing & Communication, Lancaster University, Lancaster, LA1 4WY, UK, e-mail : q.ni2@lancaster.ac.uk.

[4] Computing & Communication, Lancaster University, Lancaster, LA1 4WY, UK, e-mail : r.jiang2@lancaster.ac.uk.

The information gathered by the feedback of the nearby vehicles is often vulnerable to various security attacks, and incorrect feedback can result in higher network congestion and severe security hazards [2]. Particularly, lacking of information on end users or stakeholders will undermine the system with anonymous attacks, making it harder to pin down the responsible parties or criminals.

In the IV data sharing networks, security is always a critical issue during V2V or V2I communication. These networks requires a secure solution that can respect privacy and meanwhile enable trust [3]. Recently researchers have attempted to combine IV with the new blockchain technology [3], considering their applications based on services and smart contracts. However, blockchain itself is an anonymous technology based encryption, making it inconvenient to identify who owns the data and who are accessing the data. Obviously, lacking of knowing "who" will weaken the security of a VANET.

In this chapter, we propose a novel method, namely Biometric Blockchain (BBC), and leverage it for IV data sharing. Such a combination of biometrics and blockchain technology will enable a better secure protocol with the additive biometric information.

In our solution, biometrics are combined with the blockchain technology and provide an ID-aware credit-based trust system for transmitting reliable data over the IV data sharing platform. Here, the credit is associated with a unique biometric ID, which is then attached to the message format for the transmission in V2V or V2I communication.

The BBC based cloud storage manages the IV-targeted biometric crediting (BC) protocol, and is accessible ubiquitously by the credited end users. The IV-BC mechanism using blockchain can create the unique crypto IDs with biometric cues, self-executing digital contracts, and detailed information of IVs under the control over the BBC-based cloud [4].

Fig.1 shows the standard intelligent vehicle Information sharing environment with both V2V and V2I communication, where each vehicle in moving is dynamically routed in a peer-to-peer queue to access the internet resources via the integration of V2V and V2I communication.

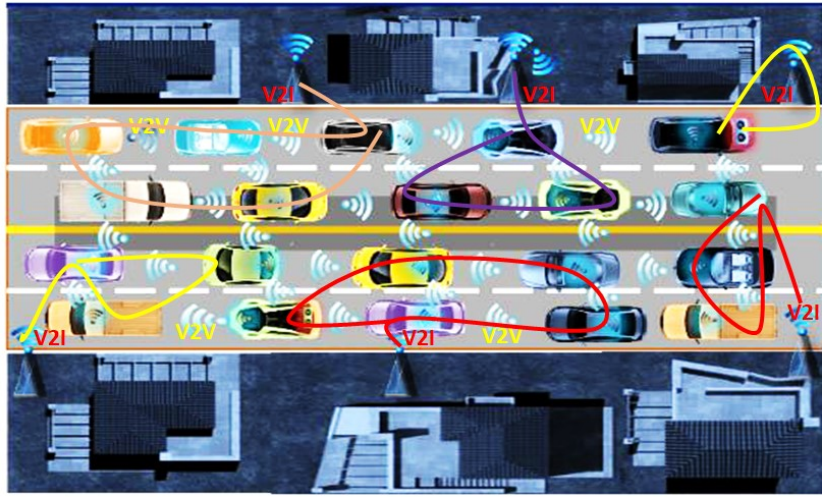

Fig. 1. Intelligent vehicles within V2I communication as well as routed peer-to-peer V2V communication.

We organize our chapter as follows. Section 2 presents the background of our work. Section 3 presents the introduction of biometric blockchain technology for Intelligent Vehicles data sharing. Section 4 presents the detailed architecture of the BBC-based IV data sharing framework. Section 5 concludes our chapter and discusses our future work on BBC-based IV data sharing.

## 2. Background of Our Work

Currently intelligent transport systems (ITS) in smart cities are mostly based on ad-hoc networks for vehicle communication. These ad-hoc networks such as DSRC, WAVE, and Cellular Network do not guarantee secure data transmission. Traditional security protocols for vehicle communication applications based on cellular and IT standard security mechanisms are often fragile to malicious attacks. They are also out of date with little suitability for real ITS applications in smart cities. So far, many researchers on ITS are advocating various new protocols to provide a robust security mechanism for IVs.

Our proposed BBC-based mechanism has many advantages. It is blockchain-based and hence very easy to be compatible to the dynamic peer-to-peer communication; the blockchain mechanism provides a natural protection to each node using its cross-validation mechanism as well as its encryption keys; with biometrics, it can easily trace down and link services to specific users or customers with lifelong historic records; and last but not least, it provides a secure and trust environment for vehicle communication with cloud-based database and ubiquitous data access in a secure way.

Our solution is based on a very simple concept of using biometric blockchain based trust environment for data sharing among *Intelligent Vehicles*. The proposed *Intelligent Vehicle Biometric Crediting* (IV-BC) protocol allows us to exploit the features of blockchain such as distributed ledgers, Merkel tree, Hash function (SHA-256), and consensus mechanism (proof of work algorithm) to build a better secure environment with the awareness of user identities. More details will be included in the following sections.

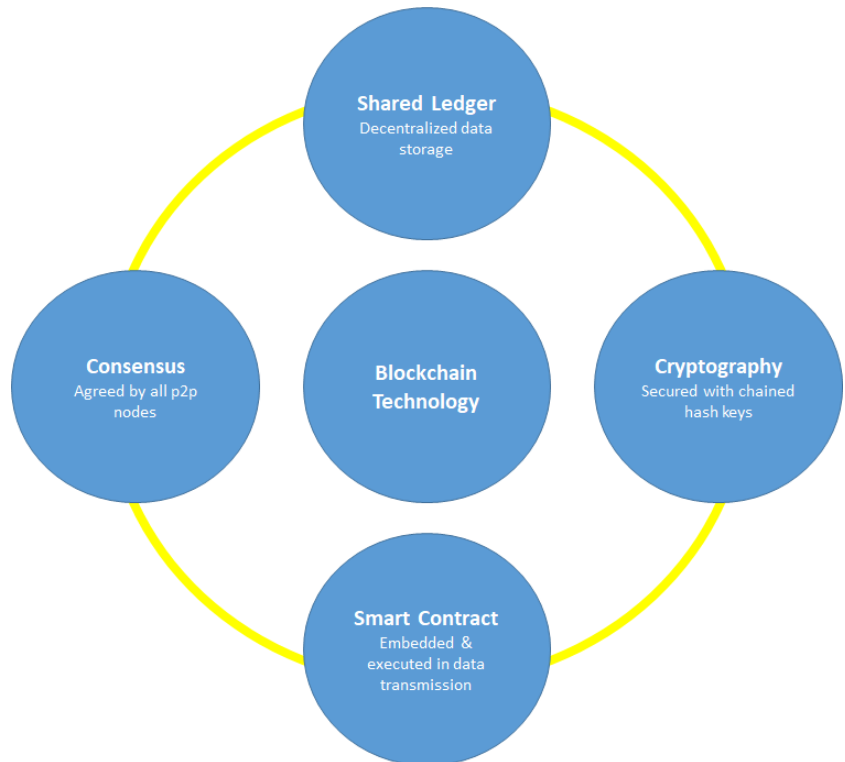

Fig. 2. Blockchain technology [6] with features such as digital signatures, shared ledger, signed blocks of transactions, cryptography.

## 3. Biometric Blockchain for Intelligent Vehicle Systems

### 3.1 Blockchain Technology

Blockchain is a new technology derived from Bitcoins. It has become a hot topic due to its high value for data security. A blockchain consists of distributed open ledgers saved at each node in a peer-to-peer network. Each-node is self-maintained. The blockchain integrity is based on a strong cryptography algorithm that validates blocks and then chains them together on transactions, making it impossible to tamper with any individual transaction without being detected. Overriding a single node in the peer-to-peer network will not hijack the whole chain and therefore it is secured from random attacks from unexpected invaders.

Fig.2 shows an overview of such blockchain technology. Within the core of blockchain, shared ledgers advocate a decentralized storage over peer-to-peer networks, the consensus protocol strengthens the data integrity by all-party validation, the smart contract mechanism enables the ledgers executed within data transmission, and a cryptography algorithm safeguards the ledgers from any malicious tampering.

### 3.2 Blockchain Technology in Vehicle

The blockchain based systems proposed in Ref.[7] employ a seven-layer conceptual model of blockchain for ITS to implement a secured, trusted and decentralized autonomous ecosystem. In Ref.[8], a self-managed blockchain-based solution is integrated with vehicular ad-hoc networks (VANET).

The introduction of blockchain technology enables the smart contract system with VANET, and promotes the combination of multiple applications at system level. For example, mandatory applications (such as traffic regulation, vehicle insurance, vehicle tax, etc.) can be easily integrated with optional applications (such as weather forecasts, traffic guidance, entertainments, e-commercial access, and other smart city functions), while blockchain technology can assist the integration of these services with trust and privacy.

With its encryption, blockchain can allow the access to peer-to-peer communication without disclosing personal information, strengthen the security in data sharing, and secure multiple communications between vehicles, etc.. In Ref.[9], a blockchain-based mechanism was proposed that will not disclose any private data information of vehicles user when providing and updating the remote wireless software and other vehicles services. In Ref.[10], the blockchain technology was exploited for securing the IV communication through the visible light and acoustic side channels. Their proposed mechanism was successfully verified via a session cryptographic key, leveraging both side-channels and blockchain public key infrastructure.

In this chapter, we propose a new technology coined as biometric blockchain and leverage it for the secure IV communication. We develop a new framework with BBC to build a secure credit-based environment for peer-to-peer communication between intelligent vehicles without interfering/disturbing other intelligent vehicles.

### 3.3 Biometric Blockchain for Intelligent Vehicles

Despite the seemingly reliable and convenient services offered by the features of blockchain technology for applications, there is a host of security concerns and issues [11-20], and its understanding is relevant to the research community, governments, vehicle industry, investors, regulators and other stakeholders.

Given the complexity and the infrastructure of the blockchain, it is not a surprise that it would have some downsides especially in sharing data among different systems or regions, and securing data for a centralised infrastructure is a challenging task since potential attacks and exploits lead to a single-point-of-contact requiring trust for this individual authority. This implies that more research efforts need to be done in order to assure that information are secured in terms of privacy, assuring that only authorized users are able to access the data.

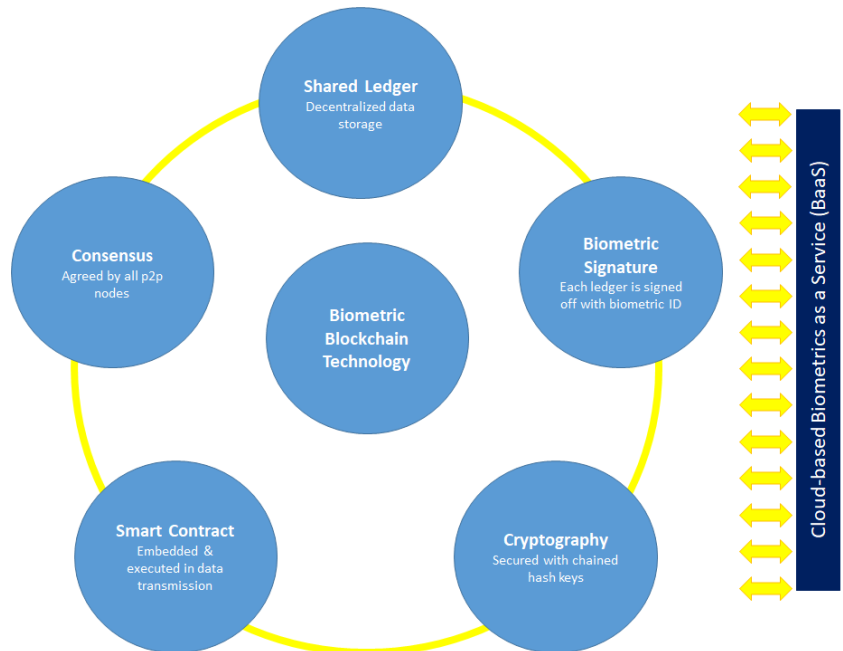

Fig.3. The proposed BBC framework for IV data sharing.

To address these challenges, recent research [20] has started to think about bring biometrics [21, 22] into blockchain with a hope to achieve better security, scalability and privacy. Based on this initiative we proposed a new blockchain framework, namely biometric blockchain (BBC) for IV data sharing.

Fig.3 shows our proposed framework, where biometrics is used as the signatures to sign off ledgers and generate unique IDs that can be associated with individuals. Cloud-based Biometrics as a Service (BaaS) is engaged to provide extra support to local IV data sharing.

The benefits of such a BBC-based framework are apparent. First, we can easily associate the activities in the IV data sharing to a specific individual or customer, and then the services can be easily tailored to specific users with their historic records over the BBC-based cloud servers.

Moreover, a personal credit-based system can be associated with user-specific services, and the security of IV data sharing is then double guaranteed by the individual credit system. Hence, BBC as proposed in Fig.3 could be a valuable solution for practical applications in IV data sharing.

## 4. BBC based Credit Environment for IV Data Sharing

In this section, we give the details of our proposed credit-based IV communication using biometric blockchain technology. Our proposed mechanism consists of three basic aspects including vehicular cloud computing (VCC), ad-hoc network enabled connected device and biometric blockchain (BBC).

Fig.4 shows a typical data-sharing environment for intelligent vehicles. Vehicles on the roads communicate with the proposed BBC-based IV data sharing platform, where each vehicle can be connected with BBC-based platform and store their IV-BC over BBC-based cloud services. More details are explained below.

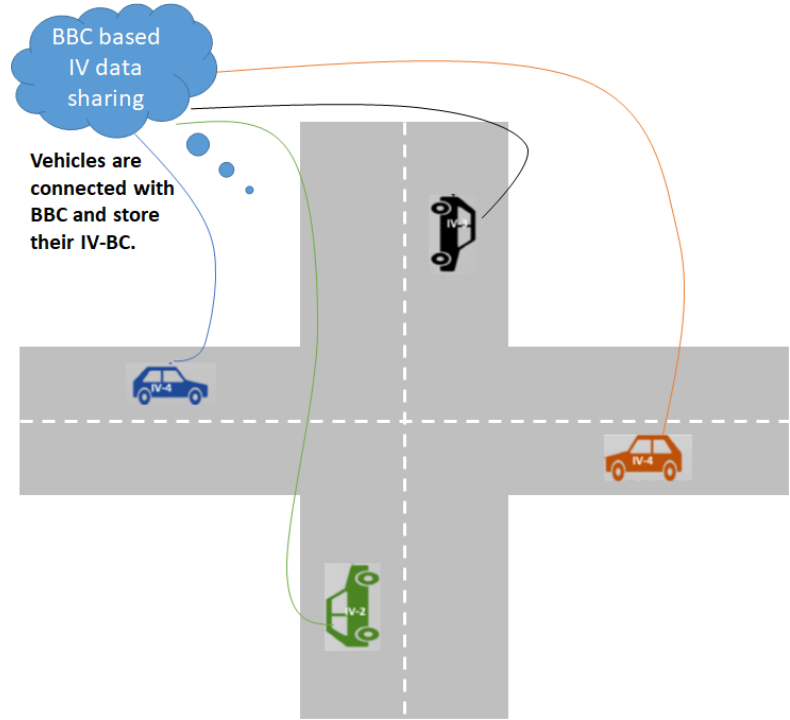

Fig.4. Proposed BBC-based intelligent vehicle communication

### 4.1 Vehicular Cloud Computing

Vehicular cloud computing (VCC) has produced a remarkable impact on traffic management and road safety by equipping intelligent vehicles with digital resources, such as cloud computing, data storage, traffic guiding, and decision-making.

VCC is a hybrid technology that vividly integrates the IV data management with cloud computing via network communication. On the other side, it faces new challenges coming from both aspects and need to handle with potential issues coming from local vehicular communication and remote cloud services.

### 4.2 Network enabled Connected Devices

While each intelligent vehicle connects itself to the infrastructure and other vehicles via an internet-enabled biometric-certificated device, such an Internet-of-Things (IoT) end-user device with biometric identification can authenticate, approve, and organize any instant communication over VANET such as Smartphone, PDA, Intelligent Vehicles, etc.

Moreover, enabled with biometric IDs, the cloud-based service can easily provide extra support based on individual information by tailoring the services to one's individual needs. Such a personalized VCC service over VANET may greatly facilitate the service providers as well as the end users.

### 4.3 BBC-Supported Intelligent Vehicles

Biometric blockchain consists of a technically unlimited number of blocks or legders that are linked together cryptographically in chronological order, signed off with individual biometric signatures. Each block consists of various transactions that are the actual data to be stored in the chain.

As shown in Fig.5, the overall system model has seven layers as a standard architecture for the intelligent communication network. Here, we briefly explain the key features of the proposed BBC-based network model, as below.

1) *Physical layer:* This layer includes the network-enabled end-user devices such as IoT devices, fingerprint ID device, face ID device, camera, GPS receiver, PDA, and other devices that are mounted on intelligent vehicles for various purpose. Mostly these devices can be involved into dynamic V2I and V2V communications with blockchain mechanism.

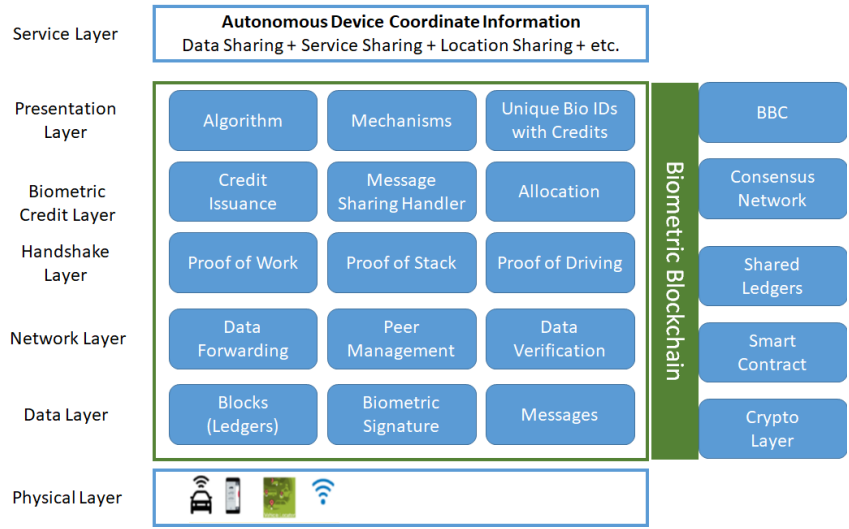

Fig.5. The proposed BBC based IV communication network framework.

2) *Data Layer:* This layer process the data blocks or ledgers with cryptography features, typically using Merkle tree and Hash algorishm to generate secured blocks. Fig.6 gives the typical structure of the blocks in a blockchain, consisting of a header that specifies the previous hash and nonce with current hash (root) following by a Merkle tree. Hash keys generated by the double SHA 256 algorithm are used to index the chained sequence.

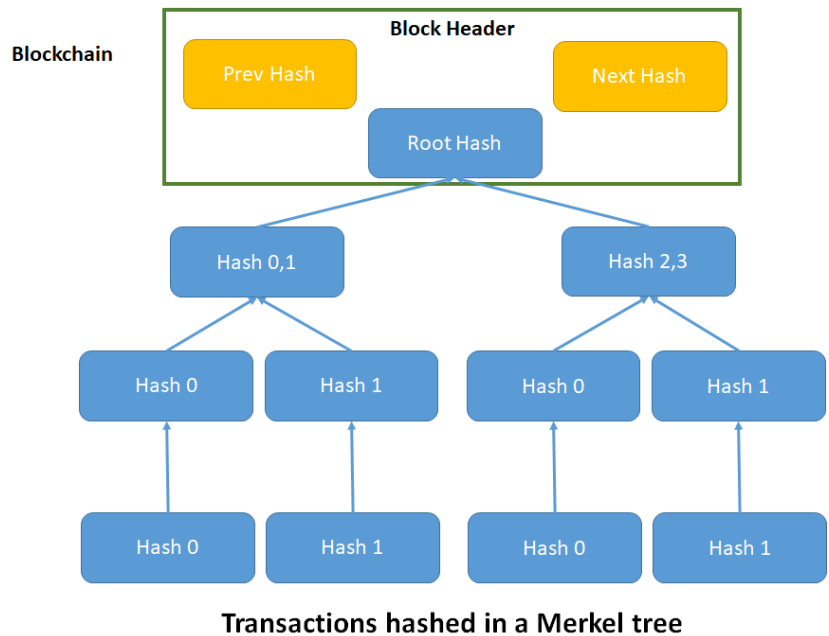

Fig.6.   The structure of blocks in a blockchain.

3) *Network Layer:* This layer handles the forwarding of data over peer-to-peer network communication. It also copes with the verification of the peer-to-peer communication, while the legality of the broadcasted message is validated and managed over the peer-to-peer interconnection between two IVs.

4) *Consensus Layer:* The consensus layer, also known as the handshake layer, provides a decentralized control in the network communication. It helps to build up the trust between unknown users in the wild communication environment, where biometric credits is exploited for such a mutual trust mechanism.

Typically in IV communication networks, *Proof-of-Driving* (PoD) can be a preferred consensus algorithm. It verifies and validates the vehicles involved in the ad-hoc communication networks. In our BBC framework, this can be also traced down to end users with their biometric credits.

5) ***Biometric credit layer:*** This layer handles with IV-BC issues, as proposed in this chapter. The proposed IV-BC protocol has a crypto data that is assigned to each vehicle, and the consensus competition will favour a vehicle with higher credits in records. The winner will be rewarded an extra credit associated with its biometric ID. The vehicle with the maximum IV-BC credit will take a lead in the communication network. Such an IV-BC credit system helps to build a reliable user-credited trust environment in the vehicles communication.

6) ***Presentation Layer:*** This layer encapsulates multiple algorithms, contracts and scripts that are provided by the vehicles and users involved in the network.

7) ***Service Layer:*** This layer contains the scenarios and user cases of IV communication system.

Currently lots of research organizations as well as startup companies are trying to implement the blockchain technology in different areas and services. In our proposed scheme, the novelty is to include personal biometric information to authorize the responsibility in a robust way and secure the environment by tracing down the history of associated key managers or users.

## 5. Privacy Issues in Biometric Blockchain

Biometrics could help secure data in blockchain. On the other side, it is also sensitive to expose the privacy of individuals. The deployment of BBC based framework will inevitably request the biometric information from individuals, such as managers or users. Hence, BBC obviously needs a privacy-protected mechanism when biometric information is collected from IV users.

Recent research has enlightened the privacy issues with a nice solution by using encrypted biometrics [21, 22, 23]. As shown in Fig.7, a set of biometric features such as faces can be encrypted and combined into a ledger as an encrypted signature. The modern biometric technology [21, 22, 23] has revealed that we do not need to decrypt these biometric information and can directly verify in the encrypted domain, making the use of biometrics much less sensitive in term of privacy concerns when it is used for blockchain technology, while BBC can identify or validate the user without knowing who the user really is.

The implementation of BBC may be associated with the online biometric verification, and can be based on multimodal biometrics [24]. Classically, feature (such as LBP or SIFT) based [25, 26] biometric verification is popular though it could be a bit time-consuming. Recently new approaches such as deep neural networks [27-30] are taking over the field of biometric verification. Such biometric verification can be carried out on cloud servers, leading to a new topic called Biometrics-as-a-Service (BaaS).

There may rise concerns on the computing resources while biometrics-as-a-service is accessed via cloud platforms, as shown in Fig.5. However, in a BBC-based platform, biometrics is verified only when it is necessary. This implies the biometric information could mostly be dormant and hence the requirement on extra computing is minimized.

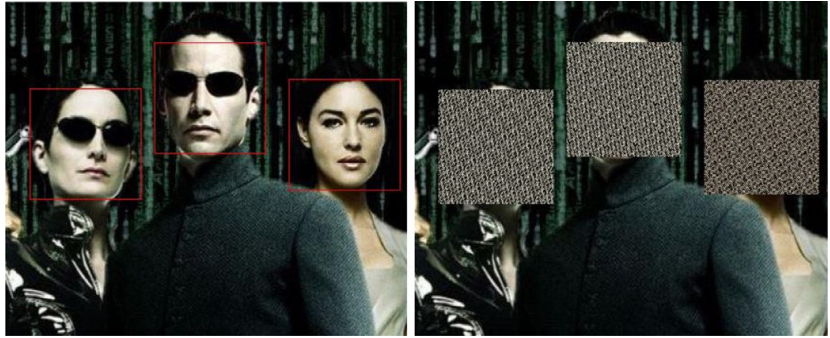

Fig.7. Privacy protection using encrypted biometrics [21].

## 6. Conclusion

In conclusion, we have presented a biometric credit based intelligent vehicle communication framework, where biometric blockchain is engaged to secure the IV data sharing and associate any activities or services with key individuals via the biometric cues. Such BBC-based IV-BC protocol can secure fast and credible communication between IVs over VANET. It also helps to detect the detailed history of the communication among IVs and end users, and a biometrics-based credit recording system can be associated to an IV with its associated individuals. Such detailed records can be further available to various user-specific individual services such as IV service providers, insurance companies, entertainment providers, and other fancy smart city functions.